\documentclass{article}
\usepackage[utf8]{inputenc}

\usepackage[inner=5cm,outer=2cm]{geometry}
\usepackage{titling}
\usepackage{authblk}
\title{\textbf{infotheory: A C++/Python package for multivariate information theoretic analysis}}
\author[1,2]{Madhavun Candadai}
\author[1,2]{Eduardo J. Izquierdo}
\affil[1]{Program in Cognitive Science, Indiana University, Bloomington}
\affil[2]{School of Informatics, Computing and Engineering, Indiana University, Bloomington}
\date{}
\usepackage{breakcites}
\usepackage[hyperindex,breaklinks]{hyperref}
\hypersetup{
    colorlinks=true,
}
\begin{document}
\maketitle
\vspace*{-1.5cm}
\abstract{This paper introduces \texttt{infotheory}: a package written in C++ and usable from Python and C++, for multivariate information theoretic analyses of discrete and continuous data. This package allows the user to study the relationship between components of a complex system simply from the data recorded during its operation, using the tools of information theory. It implements widely used measures such as entropy and mutual information, as well as more recent measures that arise from multivariate extensions to information theory, specifically Partial Information Decomposition. It provides an easy-to-use and flexible tool for use in research as well as pedgogical purposes to introduce students to information theory.\\Website: \url{http://mcandadai.com/infotheory/}\\Source: \url{https://git.io/infot}}

\section*{Background}
Information theory was first introduced by Claude Shannon in his seminal paper ``A mathematical theory of communication'' as a methodology to develop efficient coding and communication of data across noisy channels [@shannon1948]. Its rise to popularity can be primarily attributed to its ability to be applied in any domain, ranging from Economics to Neuroscience. Information theory provides a general framework to quantify stochastic properties (uncertainty in the outcome of an experiment) and relationships (mutual information that one variable provides about another) between different variables in a system of interest. It provides tools to measure these quantities in a way that is invariant to the scale of the system and allows comparison across systems.

\section*{Statement of need}
Until relatively recent times, information theory had been employed to study n-dimensional multivariate systems two variables at a time (bivariate). However, all natural systems are multivariate and a scientific inquiry into their operation requires understanding how these multiple variables interact. In a multivariate system, bivariate measures such as pairwise mutual information alone are insufficient to capture the polyadic interactions between the different variables~\cite{james2017}.
Partial Information Decomposition (PID) is an extension of Shanon information measures that allows us to study the interaction between variables in a multivariate system by decomposing the total information that multiple source variables provide about a target variable into its constituent non-negative components~\cite{williams2010}.
More specifically, in a trivariate case, the three variables can be separated into one target and two source variables. The total information that the two sources have about the target is given by the bivariate mutual information between the concatenated sources as one variable and the target. Using PID, the dependencies between the sources can be studied by decomposing this total information into the following non-negative components: information that each source uniquely provides about the target, information that they redundantly provide and the synergistic information that is only available when both sources are known.
There have been multiple approaches proposed to perform said decomposition~\cite{williams2010, griffith2014, bertschinger2014, james2018a}. Here we focus on the approach proposed by [@williams2010] primarily because this package implements PID for two and three source decomposition, and as of now, this is the only approach that guarantees non-negative decomposition for the 4 variable case (1 target and 3 sources).
Multivariate analysis allows us to ask more detailed questions such as, what is the amount of information that is uniquely provided about a target random variable by one source and not another? and what is the amount of information that is transferred from one random process X to another Y over and above Y’s own information from its past? These questions enable us to understand the interactions between different components of a complex system, thereby leading us towards an understanding of its operation given just the observed data from the system.

\section*{Features}
\texttt{infotheory} implements widely used measures such as entropy and mutual information~\cite{cover2012}, as well as more recent measures that arise from multivariate extensions to information theory.
As such, the tool has been designed to be easy to use and is ideal for pedagogical demonstrations of information theory as well as in research.
\texttt{infotheory} is open-source (\url{https://git.io/infot}) and details on how to install it and use it are available on its \href{http://mcandadai.com/infotheory/}{website}.
Here, we highlight seven key aspects of its implementation that make our package a valuable addition to any information theoretic analyses toolbox along with two existing packages, namely dit~\cite{james2018b} (focuses on discrete variables) and IDTxL~\cite{wollstadt12019} (implements an alternate approach to 3 variable PID).
First, the package is written in C++. One of the main challenges of multivariate analysis on a large, complex system is the amount of computations involved. The C++ implementation makes the package efficient.
Second, the package can be used from either C++ or Python. Python wrapping allows for ease of use, as well as compatibility with other powerful open-source libraries such as numpy.
Third, the API allows adding the data only once to then perform various analyses across different sub-spaces of the dataset cheaply.
Fourth, the data structure used to represent the random variables is sparse. This allows the package to work easily with high-dimensional data.
Fifth, to better estimate the data distribution in case of continuous variables, the package employs a kernel-based density estimation method called `averaged shifted histograms' because of its beneficial trade-off between computational and statistical efficiency~\cite{scott1985}.
Sixth, the package includes user-controllable specification of binning. This is essential for estimating distributions on hybrid systems with a mix of continuous and discrete variables.
Finally, this package implements decomposition of information in 3 as well as 4 variable systems thus making it unique among similar existing packages.

The functions implementing the above mentioned information theoretic measures have been designed to be flexibly used in alternative ways. For instance, the decomposed information components can be combined to measure transfer entropy~\cite{schreiber2000}. When dealing with time-series data, one can restructure the data such that the two sources are past values of two random variables, and the target is a future value of one of them. It has been shown that the sum of the unique information that a source provides about the target (future value) and the synergistic information from both sources is equal to the amount of information transferred from that source~\cite{williams2011}. Transfer entropy is used extensively in neuroscience to infer directed functional connections between nodes of a network (nodes can be neurons, brain regions or EEG electrodes) from recorded data~\cite{wibral2014}. Another instance of extended use of this package is to measure changes in information in time. Again, with time-series data, if the user provides all data over all time-points, then they can ask the tool to calculate all the previously discussed measures as aggregate values over time. Alternatively, the user can provide data that are only from a specific time point, calculate the information theoretic measures for that time point, and then repeat the analyses over the entire time course. Such analysis reveals how information in the variables of the system change dynamically during the course of its operation~\cite{izquierdo2015, beer2015}. Both extensions are easily accessible by reusing the existing mutual information and PID functions in the package and providing different subsets of the data accordingly.

\section*{Conclusion}
Altogether, \texttt{infotheory} provides an easy-to-use and flexible tool for performing information theoretic analyses on any multivariate dataset consisting of discrete or continuous data. Application areas are, in principle, as wide as that of information theory's - any domain that has a multivariate system and aims to study how the different components interact. We are particularly encouraged by the potential applications in neuroscience, at all scales ranging from individual neurons to brain regions to integrated brain-body-environment systems. In our group, we are currently using this package to understand the flow of information in simulated neural circuits capable of producing behavior. This tool allows us to easily analyze how different neurons of a circuit or regions in the brain are encoding information about the sensory stimulus it is receiving, the actions it is producing, or indeed about other neurons/regions within the system itself. We are using multivariate measures to analyze how different nodes in the circuit encode information uniquely, redundantly, and synergistically about a signal of interest. We are using the tool to study information dynamics of the neural circuit over time during behavior. We are also using it to infer directed functional connections between the nodes of the network. Besides its use in research, we are using this package for pedagogical purposes to introduce students to information theory. As such, we have provided a number of benchmarks and examples in the \href{http://mcandadai.com/infotheory/}{website}. We also hope to continue to extend the package in the future by, for example, implementing additional approaches to multivariate information analyses, and providing GPU-support. Finally, in the spirit of free and open-source software development, we also welcome contributions from others.

\section*{Acknowledgements}
The work in this paper was supported in part by NSF grant No. IIS-1524647. M.C. was funded by an assistantship from the Program in Cognitive Science, Indiana University, Bloomington. The authors would like to thank Randall Beer for VectorMatrix, the C++ vector libraries used in this package.

\bibliography{refs}
\bibliographystyle{apalike}

\end{document}